# Development of an Algorithm for Identifying Changes in System Dynamics from Time Series

Ferdaus Kawsar
Department of Computing
East Tennessee State University
Johnson City, TN
kawsar@etsu.edu

Mohammad Adibuzzaman
Regenstrief Center for Healthcare Engineering
Purdue University
West Lafayette, IN
madibuzz@purdue.edu

## ABSTRACT
The development of an algorithm with related mathematical concepts and supporting hypothesis for detecting changes in system dynamics from time series along with empirical analysis and theoretical justification is presented. For the method, changes in the second largest eigenvalue of Markov Chain (SLEM) or mixing rate, is observed as an indicator of the changes in system dynamics. The Markov chain is created from empirical transition probabilities of a time series. The method is developed for the application of detecting hemorrhage from arterial blood pressure in anesthetized swine. The rationale of the change in the SLEM is investigated empirically with an artificial blood pressure model and, by studying correlations with other measures such as smoothness of time series, and density of the transition probability matrix of the Markov chain. The mathematical analysis shows that the change in the SLEM is a consequence of the change in the transition probabilities between different states and reflects information about the system dynamics.

## Categories
I.2.6 [**Learning**]: Knowledge acquisition.

## General Terms
Algorithms.

## 1. INTRODUCTION
Study of dynamical systems describe the temporal evolution of a system. Different measures such as auto-correlation, eigenvalues of Markov chain, stationary probability and reconstructed phase space (RPS) have been studied to identify this temporal evolution [1] [2] [3] [4]. Markov chain analysis also provides a way to describe the nature of dynamics. We present the development of an algorithm that utilizes the change in properties of a Markov chain constructed from time series data, namely second largest eigenvalue of the Markov chain (SLEM), as an indicator of the changes in dynamics of the system from which the time series was generated.

Markov chains have recently become a standard tool in many applications; especially through Markov chain Monte Carlo (MCMC) methods and hidden Markov models (HMM), both of which have been applied in a large number of settings [5]. For our approach, one of the properties of convergence to a limit distribution, second largest eigenvalue of a Markov chain, has been proposed as a feature of the chain, similar to MCMC methods. We present this novel index for identifying changes in system dynamics. The development of the algorithm is described with the hypothesis that dynamical properties of a Markov chain constructed from arterial blood pressure would change as an animal undergoes hemorrhage. The hypothesis for the change in the SLEM is constructed and explained with the example of discrete logistic equation. The parameters of the algorithm were set by empirical experiments. This manuscript focuses on the validity of the method with empirical and mathematical analysis. Empirical analysis compares the index with other approaches such as smoothness measures of time series and properties of a matrix such as density and self-transition probability. The empirical evidence (high correlation with density and self-transition probability) led to the theoretical justification of the change in the SLEM of the transition probability matrix with Gershgorin circle theorem [6]. The approach is also compared to another method that captures dynamical change. To investigate the change in the SLEM with regards to signal properties, we changed the parameters of an artificial blood pressure model and observed corresponding change in the SLEM. The results show that the change in the SLEM is explained mostly by the structural changes in the transition probability matrix. The resulting index, SLEM, has been found to have high correlation with shock index (median correlation coefficient of -0.95), which is the ratio between heart rate and systolic blood pressure and is widely used in clinical setting for predicting likelihood of mortality [7]. The significance of this finding is that blood pressure signal alone would suffice to predict the instability of a subject without the need for capturing heart rate. Furthermore, this algorithm can be explored to identify dynamical change from time series data where similar change is expected. To summarize, the contribution of this manuscript includes,

- Empirical development approach of the algorithm for detecting dynamical change with supporting hypothesis.
- Empirical analysis of the algorithm with theoretical justification.
- Comparison of the algorithm with another method for detecting dynamical change.
- Finding the significance of the index with regards to signal properties with an artificial blood pressure model.

The technical discussion of this article begins with a brief introduction of related mathematical concepts and definitions.

## 2. BACKGROUND
This section illustrates the basic mathematical background and concepts related to the algorithm development such as eigenvalues,

Markov chain, and properties of Markov chain. We also define the quantities we investigated as there is variability in the definitions of these measures. We briefly introduce the readers to the creation of random paths from Markov chain and, also to the approximation of the chain, given a transition probability matrix.

## 2.1 Eigenvalues

If a nonzero vector (eigenvector of a matrix) is multiplied by a matrix, the eigenvalue of a matrix is the constant by which amount the vector is scaled up or scaled down. For example,

$$B \times v = \lambda \times v$$

Here $v$ *is the right eigenvector* of the matrix $B$ and, $\lambda$ is the eigenvalue for the corresponding eigenvector. Geometrically speaking, the definition states that when the eigenvector is multiplied by the matrix, if there is an eigenvector, the resultant is not rotated (as this is equivalent to multiplying with a scalar). This idea can be extended for any vector since any vector can be represented by a basis whose reference vectors are eigenvectors (Figure 1) of the transition matrix, in the case that there is a complete set of eigenvectors and the eigenvalues are distinct.

## 2.2 Markov Chain

Markov chains were first introduced as an extension of work related to the Law of Large Numbers to dependent events by Andrei Andreyevich Markov [8]. The basic idea of Markov chain is that the next value of a sequence depends on previous values, unlike a sequence of independent random numbers. This dependence can be of different orders. The simplest Markov assumption, the first order Markov chain, is that the next value of a sequence depends only on the current state. Markov chains can be discrete or continuous. A discrete Markov chain is a mathematical process that transits from state to state. It consists of a countable set of states S, and corresponding transition probability matrices, $P(n)$, that describes the state transition probability at time $n$. If there are $N$ states, the matrices $P(n)$ will be $NxN$. The $i, j^{th}$ entry in the transition matrix, $p_{i,j}(n)$, is the conditional probability that the next transition will be to state $j$ if the system is in state $i$ at stage $n$. A generalized Markov chain can be of order '$n$', indicating the transition probability from previous '$n$' time steps. In its simplistic form, we assume the transition probability matrix $P_{i,j}(n)$ is independent of the time '$n$'. In that case, the chain is first order Markov chain, and only one transition probability matrix is required to describe the chain.

## 2.3 Eigenvalues of Markov Chain

For any iterative method that involves multiplication of a vector by a matrix in repeated form, the component related to the maximum eigenvalue, if there is one, dominates in the long run. Consequently, eigenvectors of a Markov chain describes the probability distribution of the system of being in different states after many steps. Multiplying the transition probability matrix many times with some initial state probabilities, is equivalent to multiplying the state probabilities many times with the eigenvalues. For example, *any vector* $v$ can be represented by a linear combination of the eigenvectors $v_1$ and $v_2$ ($v = av_1 + bv_2$) of a matrix $B$ (Figure 1).

$$B^i \times v = B^i \times (av_1 + bv_2) = B^i \times av_1 + B^i \times bv_2$$
$$= \lambda_1^i \times av_1 + \lambda_2^i \times bv_2$$

As a result, the eigenvalues that are less than 1 in magnitude soon make the system converge to a *limit distribution* or *steady state*, which corresponds to the eigenvector corresponding to the eigenvalue, 1. Because of the construction, for any transition probability matrix of a Markov chain, there exists a unique eigenvector, $\pi$ for which the eigenvalue is 1 (if 1 is a simple eigenvalue). This is called the *limit distribution* or *steady state distribution*. Furthermore, except for rare or constructed situations, all other eigenvalues are less than one in modulus. Consequently, *the second largest eigenvalue of the Markov chain* (SLEM) essentially determines how fast any initial state would converge to the limit distribution. The smaller the SLEM is, the faster the convergence would be. After $n$ steps, where $n$ is sufficiently large, the stationary distribution contains only the eigenvector corresponding to the eigenvalue 1. This property is the basis for many methods such as the PageRank algorithm [9] and principal component analysis [10].

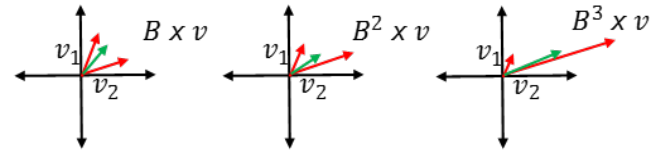

**Figure 1: Repeated multiplication of the vector $v$ with the matrix B with two eigenvalues. $v$ is represented as a linear combination of the eigenvectors ($v_1, v_2$) of the matrix B. Here the eigenvalue $\lambda_2 > \lambda_1$, and as a result, the resultant vector aligns with the eigenvector $v_2$, corresponding to eigenvalue $\lambda_2$.**

## 2.4 Sample Path (Time Series) from Markov Chain (Transition Probability Matrix)

A sample path can be simulated using the transition probability matrix of a Markov chain with a random number generator [1]. This can be achieved by at first creating a matrix that has the cumulative probability for each row of the transition probability matrix. A uniform random number is generated, and compared with the cumulative transition probability of that row or the state. Depending on the difference between these two numbers, the system moves to a new state or not.

## 2.5 Markov Chain (Transition Probability Matrix) from Sample Path (Time Series)

Conversely, Markov chain or the transition probability matrix of a Markov chain can be computed from a sample path. This is achieved by constructing a matrix of the relative frequency of observed transitions. The relative frequencies can be computed by measuring each transition and having a histogram for state to state transition. This approximation of the chain is one phase of our approach (Algorithm 1). The approximation usually gets more accurate as the number of samples increases [1].

## 2.6 Definitions

Here, we define the quantities that have been used in the manuscript for the empirical analysis of our approach.

### 2.6.1 Smoothness of time series

The smoothness of a time series is measured as the summation and variance of the $1^{st}$ and $2^{nd}$ order derivatives.

### 2.6.2 Density of a matrix

Density of a matrix is defined as the number of non-zero elements of the matrix divided by the total number of elements. For a $NxN$ square matrix,

**Table 1. A summary of the notations for measures of smoothness**

| | |
|---|---|
| $S_1$ | Summation of the 1st derivatives of a time series |
| $S_2$ | Summation of the 2nd derivatives of a time series |
| $V_1$ | Variance of the 1st derivatives of a time series |
| $V_2$ | Variance of the 2nd derivatives of a time series |

$$Density = \frac{Number\ of\ Non\ Zero\ Elements}{N \times N}$$

*2.6.3 Self-transition probability of a Markov chain*

For a Markov chain, the summation of the probabilities that the chain would stay at the current state, is defined as the self-transition probability. For a Markov chain with N states, if the transition probability matrix is P,

$$Self-transition\ Probability = \sum_{i}^{N} P_{ii}$$

# 3. HYPOTHESIS CONSTRUCTION: SIMILAR DYNAMICAL PROPERTIES WITH A SIMPLE EXAMPLE

For the development of the algorithm, we examined the dynamical properties of a Markov chain and hypothesized that one or more of those properties of Markov chain constructed from arterial blood pressure would change after the start of hemorrhage in an animal. For the selection of these properties of the Markov chain and hypothesis generation, we introduce '*similar dynamical properties*' with respect to Markov chain and transition probability matrix with the example of discrete logistic equation. These measures of similar dynamics contributed to the formation of the hypothesis of the change in the SLEM as an indicator of dynamical change. Different measures can be used for defining similar dynamics [1].

- Eigenvalues of the transition matrices away from zero show a similar blend of positive, negative, and complex values. The limit distributions have the same number of nonzero values in corresponding states and close as points in $\mathbb{R}^N$.
- Autocorrelation function (ACF) show similar patterns of correlation.
- The spectral radius of the difference of the two matrices is small. This is defined as the "norm" of a matrix A, $norm(A) = (\max(eigenvalue(A'A)))^{1/2}$
- Phase plots show structural similarity for certain time lag and dimension [2].

These concepts of similar dynamics are explained with the example of discrete logistic equation,

$$x_{n+1} = \mu x_n (1 - x_n) \quad (1)$$

For our example, μ=3.8 is used to generate a sample path of length 1000. Then, noise is added to this sample path in two different ways. First, by adding random noise with each observation with a zero mean and standard deviation 0.1; noise is added to each $x$. This is to simulate measurement error, and does not change dynamic properties.

$$x_{n+1} = \mu x_n (1 - x_n)$$

$$z_n = x_n + \epsilon_n \quad (2)$$

To introduce dynamic change, we add noise in the dynamics of the series with the following equation, with a dynamic noise of mean zero and standard deviation 0.02.

$$x_{n+1} = \mu x_n (1 - x_n) + \epsilon_n \quad (3)$$

The difference between (2) and (3) is that the later propagates the added noise to each successive $x_n$. A thousand samples are generated for each equation, which are shown in Figure 2(a), 2(b) and 2(c). For each of the time series, we create an empirical Markov chain; and the dynamical properties, such as, limit distribution, eigenvalues of the transition probability matrix of the Markov chain, auto correlation function, spectral radius of the difference matrix from the baseline, and the phase plot, are plotted (Figure 2 shows three of these properties). From observation, we find that for equation (1) and (2), i.e., for the measurement error (column (1) and (2), Figure 2), there is no significant change between the SLEM. Phase plots with time lag 1 also shows similar distribution. However, there is a subtle change in the dynamical properties when dynamic error is considered (between equation (1) and (3), column (1) and (3), Figure 2). This is observed in complex eigenvalues in the second largest eigenvalue of the Markov chain (SLEM), and the distribution of the points in the phase plot. Other dynamical properties such as the distribution of steady state probabilities (limit distribution), autocorrelation function and spectral radius showed similar behavior. Due to the page limitation, those figures are not shown in this manuscript. Similar to this discrete logistic equation, we hypothesized, for the dynamical event of hemorrhage in animals, there would be change in some of these properties of Markov chain constructed from the arterial blood pressure data.

# 4. OUR APPROACH

The algorithm is empirically developed by investigating if there is any identifiable change in the dynamical properties of a Markov chain constructed from time series. The time series investigated here is the arterial blood pressure of swine undergoing hemorrhage. The event for dynamical change is the start of hemorrhage in the swine. The hypothesis was that there would be some observable and repeatable change in the dynamic properties of the Markov chain constructed from arterial blood pressure indicative of hemorrhage. The algorithm, proposed in [11], is developed in two phases: first, construction of the transition probability matrix with a fixed number of states from time series (Algorithm 1) and second, detection of change in the distribution of the SLEM after hemorrhage (Algorithm 2).

We selected the problem of hemorrhage detection as our research goal as detection of hemorrhage in a timely manner is important in many situations for adequate intervention. 40% of deaths after a traumatic injury in the United States are caused by hemorrhage [12]. If untreated, hemorrhage may cause hemodynamic instability, inadequate tissue perfusion or even death [13]. Due to the nature of hemodynamic collapse, it is important to identify hemorrhage as early as possible. But, because of the action of the sympathetic and parasympathetic nervous system, traditional vital signs may remain unchanged until later stage of hemorrhage, limiting the detection of hemorrhage. As a result, small decrease in detection time might end up being the difference between life and death.

Changes in blood pressure and heart rate as a consequence of hemorrhage have been reported in the literature [14] [15].

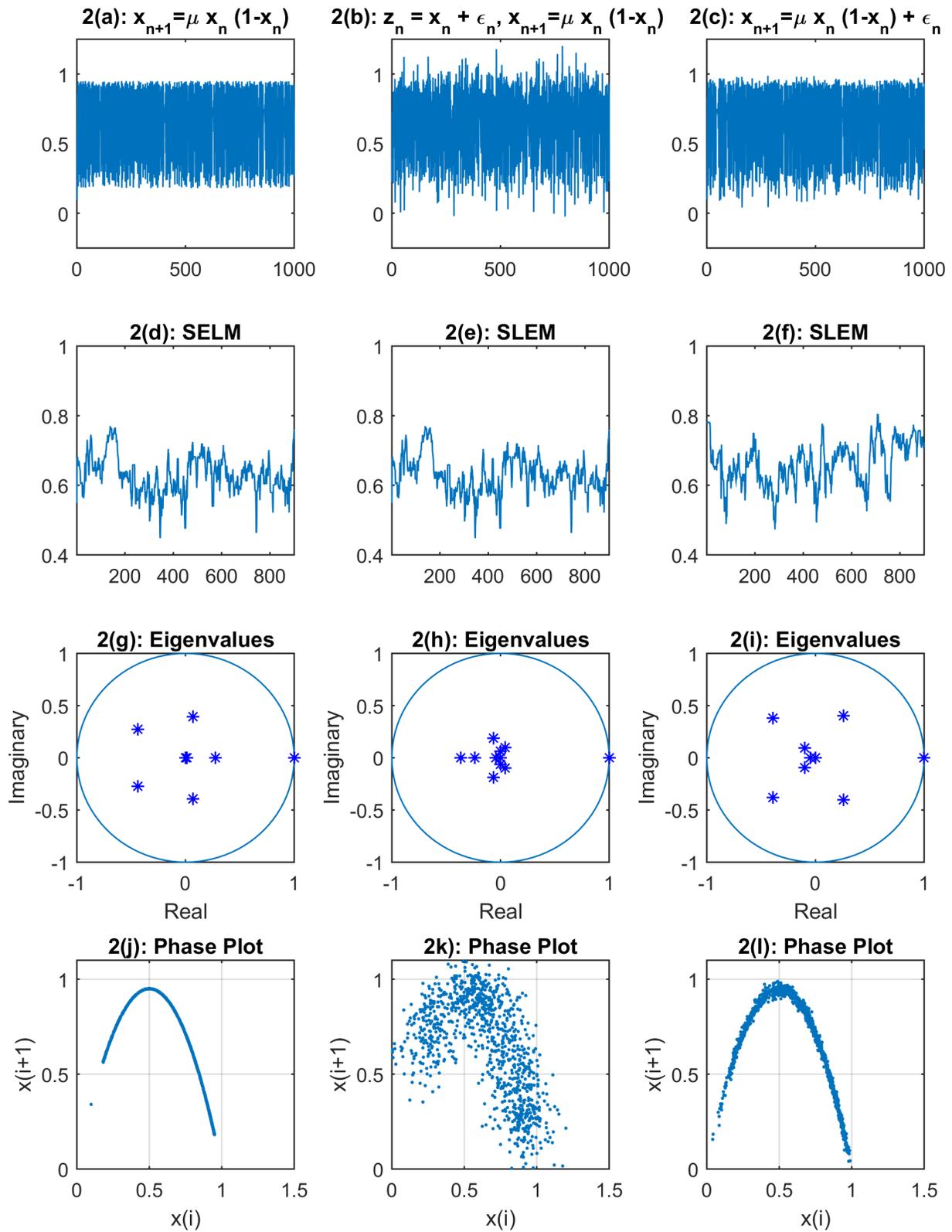

Figure 2: The three time series for discrete logistic equation with different ways of noise addition. 2(a) is the series with no noise; 2(b) is noise added to each measurement and 2(c) represents the series with dynamic noise added. The corresponding subplots show the different dynamic properties of the data including limit distribution of the Markov chain, eigenvalues of the transition probability matrix, and phase plot in 2-dimension.

**Algorithm 1:** Markov chain transition matrix from time series

**Data:** Time Series Data (d), Number of States (m)
**Result:** $m \times m$ Markov chain transition matrix (ProbNormalized)
min = minimum of the data (d)
max = maximum of the data(d)
width = (max-min)/m
\\Count the bin number for each data point
**for** $i=1$: length of the data **do**
    binNum(i)=floor((d(i)-min)/width)+1
    **if** $binNum(i)==m+1$ **then**
        binNum(i)=m
\\Compute the number of transitions from each bin to any other bin
**for** $i=1$:length of the data -1 **do**
    D(binNum(i),binNum(i+1))= D(binNum(i),binNum(i+1))+1
**for** $j=1$:m **do**
    **if** $sum(D(j,:)>0$ **then**
        ProbNormalized(j, :)=D(j,:)/sum(D(j,:))
    **else**
        ProbNormalized(j, :)=0
**return** ProbNormalized

**Algorithm 2:** Change point detction in the distribution of SLEM

**Data:** Second largest eigenvalue for each window (SLEM)
**Result:** Indicator for change detection(detectionFlag); If successful, index of first alarm(indexFirstAlarm)
baselineWindowSize = 75  \\75 seconds with 20 second overlap
downSampleRate = 4
$\alpha = 5$
nextWindowSize = 4
startIndex = baselineWindowSize+1
uncorrelatedSLEM = **downsample**(SLEM,downSampleRate)
95thPercentile = 95th percentile for uncorrelatedSLEM
alertFlag=0
**for** $i=1$:$floor(\frac{length_{(uncorrelatedSLEM)} - baselineWindowSize}{nextWindowSize} - 1)$ **do**
    nextSample = (uncorrelatedSLEM(startIndex:startIndex+nextWindowSize))
    **if** each of nextSample is less than 95thPercentile **then**
        **if** detectionFlag==0 **then**
            indexFirstAlarm = startIndex
            detectionFlag = 1
    startIndex = startIndex+1;
**return** detectionFlag, indexFirstAlarm;

In one example, morphological change as well as increase in heart beat in arterial blood pressure are observed for a swine between before and after hemorrhage across a two-second time window (Figure 3). Similar changes have also been suggested in the literature for human after simulated hemorrhage with lower body negative pressure [16]. These approaches have their own limitations; for example, heart rate increase can be an indicator of anxiety instead of hemorrhage [17]. Consequently, there is a need for more sensitive technique for hemorrhage detection as well as the need for the decrease in the required time for detecting hemorrhage. To develop the algorithm, we investigated if there is any change in features of the Markov chain constructed from arterial blood pressure due to the dynamical event of hemorrhage. Candidates for indicators of dynamic changes included: a) eigenvalues, b) limit distribution, c) autocorrelation, d) spectral radius, e) SLEM and f) phase plot, as explained by the example of discrete logistic equation. Data were obtained from the University of Texas Medical Branch at Galveston from pigs undergoing hemorrhage and the protocol was approved by the University of Texas Medical Branch Institutional Animal Care and Use Committee (IACUC).

Algorithm 1 shows the construction of the Markov chain transition probability matrix from time series. In our case, the time series is the arterial blood pressure data from pigs undergoing hemorrhage. A moving average filter (2000 samples) was used to remove the effect of the change in mean arterial blood pressure. After the construction of the chain, the Blue Meadow cluster from the FDA Scientific Computing Laboratory with Octave parallel computing was used to test the hypotheses with varying window sizes (5-30) and number of states (5-30) for each of the dynamic properties. The Markov chain was constructed for each window. Visual images were obtained and the change in the dynamic properties were observed after the hemorrhage for each swine. The resultant images showed only the change in the SLEM consistently across all the hemorrhages. Figure 4 shows the change in the SLEM after hemorrhage for all the pigs (N=7). However, the magnitude of the changes varied across different swine. Algorithm 2 shows the method to identify changes in the SLEM. The parameter values are set for the best results out of different combinations. A detail of the experiment is published in [11]. The SLEM showed to have changed from baseline estimate across all animals (N=7, median time of 13 minutes).

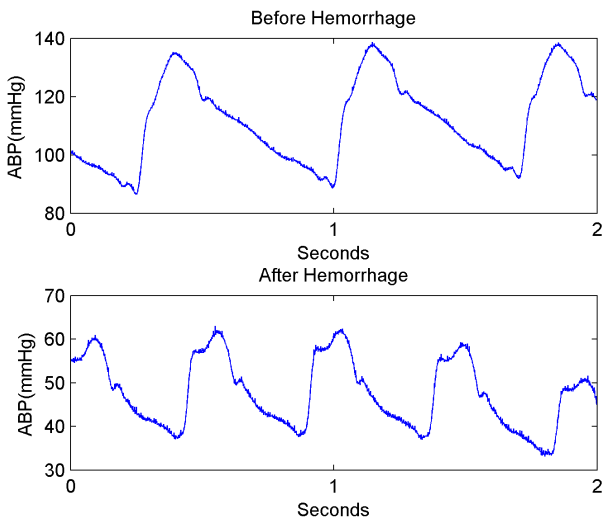

Figure 3. Example of arterial blood pressure waveform morphology for a swine before and after hemorrhage within two seconds.

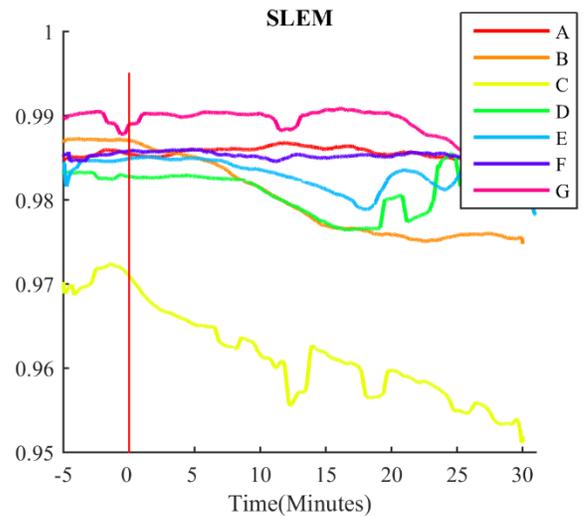

Figure 4. The second largest eigenvalue of a Markov chain (SLEM) for 7 swine. Each color represents a different pig. The red Vertical line represents starting point of hemorrhage.

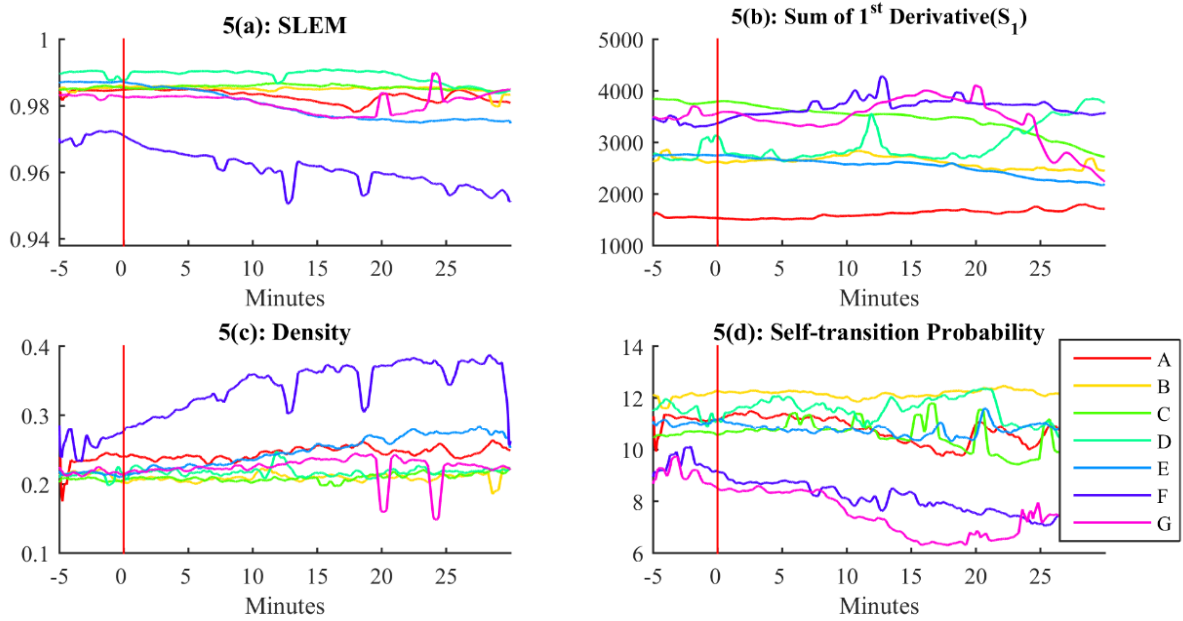

**Figure 5.** The change in different measures such as smoothness, density and stationary probability of the transition probability matrix along with the changes in s the second largest eigenvalue (a) during hemorrhage. Density (c) and stationary probability (d) have high correlation with the second largest eigenvalue, whereas the smoothness measures (1st derivative (b)) does not. The red vertical lines indicate start of hemorrhage and each color represents a different animal.

## 5. EMPIRICAL ANALYSIS

The rest of the manuscript focuses on the understanding of the rationale for the change in the SLEM with empirical and mathematical analysis. The nature of the changes in the morphology such as increase in heart rate and decrease in heart rate variability [14] suggests that hemorrhage causes change in the variability of the arterial blood pressure. Consequently, other measures of smoothness and variability in the signal and the correlation between the SLEM and those measures were examined. The understanding of the change in SLEM is presented with two perspectives: 1) correlation to measures of *smoothness* or *roughness* (different derivatives of the signal), and 2) correlation with changes in the structural properties, e.g., density or self-transition probability, of the Markov chain *transition probability matrix*.

### 5.1 Smoothness Measures

Observation from the morphological change suggests that the smoothness or roughness of the blood pressure changes with progressive hemorrhage. Heart rate variability, a measure that could be computed from arterial blood pressure, has also been reported to change in the literature due to hemorrhage [14]. In mathematics, different orders of derivatives are used as a measurement of the smoothness of the graph [18]. We measured the smoothness for each window (20 seconds, 2000 samples, same window that was used to calculate the SLEM) as the summation and variance of the 1st and 2nd order derivatives. They are listed as $S_1, S_2, V_1$ and $V_2$. Table 1 summarizes these notations. While the smoothness changed significantly in most cases, the change was not consistent across all the animals. Different pigs responded differently; probably because of variability in their autonomic nervous system response. Some pigs showed an increase in the smoothness measure whereas some pigs showed a decrease. Figure 5(b) shows the smoothness as the sum of the 1st derivatives for each time window before and after the hemorrhage. Figure 6 shows the box plot for the correlation coefficients with the SLEM and different measures of smoothness.

### 5.2 Properties of Transition Probability Matrix

The minimization of the SLEM has been studied in Mathematics as a convex optimization problem [19]. In this area, the goal is to assign the transition probabilities such that the SLEM is minimized. A common heuristic is to use the degree or density of a Markov chain as an indicator of the convergence rate. A higher degree indicates more transitions between the states of the chain and causes faster convergence without any 'bottleneck'. In other words, higher probability for self-transition would limit the chain to a slow convergence rate. We investigated whether the degree of the transition probability increases or the self-transition probability decreases with progressive hemorrhage. Both of these phenomena were observed for all the animals (Figure 5(c) and 5(d)). The number of transitions between the states was measured in terms of 'density'. Consequently, the correlation coefficient between the SLEM and both of these parameters were found to be high (Figure 6).

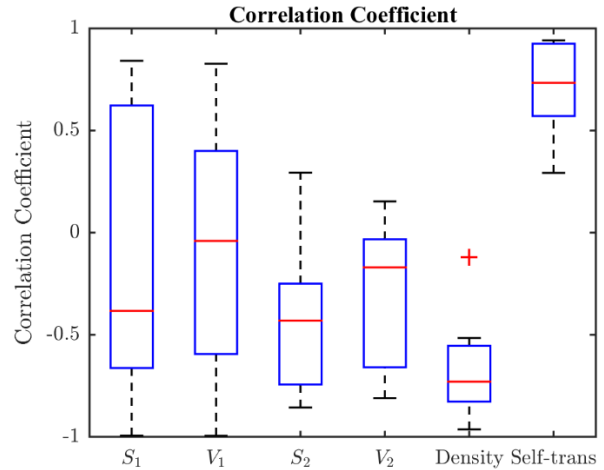

**Figure 6.** Box plot for correlation coefficient with the SLEM and other measures. The notations on the X axis are listed in Table 1. Each box plot represents the correlation coefficient for all the animals (N=7).

## 6. THEORETICAL JUSTIFICATION

The empirical analysis shows that the change in the SLEM is highly correlated with the change in the structural properties of the transition probability matrix. It has a negative correlation with density and positive correlation with self-transition probability (Figure 6). For the theoretical justification of the algorithm, we considered this empirical evidence of high correlation between the SLEM and self-transition probability which led to the Gershgorin circle theorem that describes the relationship between eigenvalues and the structural properties of the transition matrix [6].

**Gershgorin Circle Theorem:** Let $B$ be an arbitrary matrix. Then the eigenvalues $\lambda$ of $B$ are located in the union of the n disks,

$$|\lambda - b_{kk}| \leq \sum_{\substack{j=1 \\ j \neq k}}^{N} |b_{kj}| \quad (4)\,[6]$$

Where $b_{kk}$ is the diagonal element (self-transition probability) and $b_{kj}$ is the non-diagonal elements for each row (related to density) of the matrix.

For the analysis, from equation (4), we get,

$$\lambda - b_{kk} \leq \sum_{\substack{j=1 \\ j \neq k}}^{N} |b_{kj}|$$

Or,

$$\lambda \leq \sum_{\substack{j=1 \\ j \neq k}}^{N} |b_{kj}| + b_{kk} \quad (5)$$

And,

$$-(\lambda - b_{kk}) \leq \sum_{\substack{j=1 \\ j \neq k}}^{N} |b_{kj}|$$

Or,

$$\lambda \geq \sum_{\substack{j=1 \\ j \neq k}}^{N} -|b_{kj}| + b_{kk} \quad (6)$$

From (5) & (6),

$$\sum_{\substack{j=1 \\ j \neq k}}^{N} -|b_{kj}| + b_{kk} \leq \lambda \leq \sum_{\substack{j=1 \\ j \neq k}}^{N} |b_{kj}| + b_{kk}$$

The construction of the Markov chain is such that the summation of each row is 1. As a result,

$$\sum_{\substack{j=1 \\ j \neq k}}^{N} -|b_{kj}| + b_{kk} \leq \lambda \leq 1$$

As more states transition to the other states, the self-transition probability $b_{kk}$ decreases and $\sum_{\substack{j=1 \\ j \neq k}}^{N} |b_{kj}|$ increases, resulting the decrease in the lower bound of this inequality. This is what we observed in the empirical analysis after the hemorrhage (Section 5.2). We conclude, the lower bound for the SLEM decreases, as captured by the algorithm.

## 7. COMPARISON WITH AN APPROACH FOR DYNAMIC CHANGE DETECTION

In this section, we compare the time required for hemorrhage detection between our approach and another method for the detection of dynamical change using reconstructed phase space (RPS). Dynamical systems have been studied in non-linear systems [20]. At any time, a state of a dynamical system can be represented by a set of real numbers or a vector. The next state can be determined by an evolution rule, defined as the dynamics of the system, which can be deterministic or stochastic. A state space or phase space is the set of all possible states. Takens [3] and Sauer et al. [4] showed that a time series of observations from a single state variable (in our case, blood pressure) can be used to reconstruct a space or time delay embedding topologically equivalent to the original system. The state of a time series of dimension $d$ and time lag $\tau$ is defined as a row vector,

$$X_n = [x_{n-(d-1)\tau} \ldots \ldots x_{n-\tau}\, x_n]$$

Where $n = (1 + (d-1)\tau) \ldots \ldots N$. Each row vector $X_n$ is a point in RPS. The dimension $d$ can be approximated using the nearest neighbor algorithm. For selection of time lag $\tau$, a common heuristic used in the literature is the first minimum of the auto mutual information function [21]. Each state of the dynamical system is a point containing all the information to compute any future state following t. This might be a more robust system which does not hold the Markov assumption, often not true in time series data. Figure 7 shows that in the reconstructive phase space with $d = 3$ and $\tau = 1$ the distribution of the state space is different between before and after hemorrhage for the arterial blood pressure. Gaussian mixture models (GMM) can be used to differentiate the distributions in the reconstructed phase space [2]. The method works in three steps: first, creating the RPS with appropriate dimension and time; second, learning the GMM for before hemorrhage. Third, signal classification with an empirical threshold. If the probability of the signal being in the GMM model trained using the data before hemorrhage is significantly low (lower than the empirical threshold value), the model predicts a hemorrhage. Table 2 shows the comparison of the time required for hemorrhage detection using our approach and Gaussian mixture models of the reconstructive phase space with $d=3$ and $\tau=1$. We found that our approach works better as this method failed to detect hemorrhage in some cases (animals F and G).

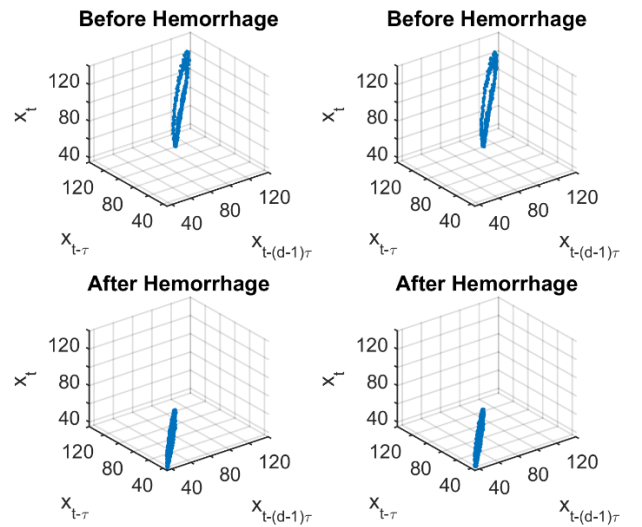

**Figure 7. Phase plot for two sample signal of length 500 (5 seconds with 100 Hz) before hemorrhage and two sample signal after he hemorrhage from a pig ($d = 3$ and $\tau = 1$).**

**Table 2. Comparison of detection time of hemorrhage using RPS and SLEM**

| Animal | Time till change detected using RPS (minute) | Time till change detected using change in SLEM (minute) |
|---|---|---|
| A | 19 | 25 |
| B | 13 | 2 |
| C | 5 | 7 |
| D | 19 | 13 |
| E | 13 | 17 |
| F | NA | 33 |
| G | NA | 12 |
| Group Statistics | | |
| Median | 13 | 13 |
| Min | 5 | 2 |
| Max | 19 | 33 |

## 8. EFFECT OF SIGNAL PROPERTIES ON THE CHANGE IN THE SLEM

To understand the relationship between the change in the SLEM and the properties of a signal, different signal properties of blood pressure in a dynamical model were changed, and the corresponding changes in the SLEM were observed. A dynamical model developed by McSharry et al. was used for generation of artificial blood pressure [22] [23]. The model uses a three dimensional state space for simulating the trajectory of each heartbeat. Quasi-periodicity of the blood pressure signal is created by the movement around an attracting limit cycle of unit radius in the (x, y)-plane. One heart beat is represented by each revolution around the unit circle. The motion in the z-direction imputes inter beat variation. Different extrema in the z direction represents the distinct points of an ECG signal, namely P, Q, R, S and T. The model uses a specific set of parameters for creating ECG or blood pressure signals. Different parameters of the model can be changed to simulate progressive hemorrhage. Morphological change similar to hemorrhage can be introduced as well. The dynamical equations of motion are given by a set of three ordinary differential equations,

$$\dot{x} = \alpha x - \omega y$$
$$\dot{y} = \alpha y + \omega x$$
$$\dot{z} = -\sum_{i \in \{P,Q,R,S,T\}} a_i \Delta\theta_i \exp\left(\frac{\theta_i^2}{2b_i^2}\right) - (z - z_0) \quad (7)$$

Here,

$$\alpha = 1 - \sqrt{x^2 + y^2}$$
$$\Delta\theta_i = (\theta - \theta_i) \mod 2\pi$$
$$\theta = atan2(y, x)$$

And ω is the angular velocity of the trajectory.

To simulate hemorrhage with the artificial blood pressure model, we investigated the parameters that change due to hemorrhage. For example, due to the loss of blood volume, sympathetic nervous system becomes dominant with the withdrawal of vagal tone [15]. Consequently, heart rate increases to compensate for the loss of blood pressure. Based on this, we increased the heart rate in the dynamical model, and we observed a decrease in the SLEM (Figure 8). Table 3 shows the list of the parameters of the model for which the change in the SLEM was observed. Each parameter was changed independently of the others. The change in the magnitude of the SLEM for a unit change in the parameter is also shown for the variables that show a high correlation with the change in the SLEM. For example, heart rate variability (HRV) has a high positive correlation (0.83) with the SLEM. For a unit change in HRV, a magnitude of 72.37x10$^{-5}$ change in the SLEM was observed. HRV was measured as standard deviation of heart rate.

**Table 3. Correlation coefficients with the SELM and parameters of an artificial blood pressure model along with unit change in the SLEM for highly correlated parameters.**

| Parameter [Default value] | Correlation coefficient with mean SLEM |
|---|---|
| sfecg: ECG sampling frequency [256 Hertz] | 0.94 $\frac{\Delta\lambda}{\Delta sfecg}$=1.74x10$^{-5}$ |
| hrmean: Mean heart rate [60 bpm] | -0.96 $\frac{\Delta\lambda}{\Delta hr}$=-4.18x10$^{-5}$ |
| hrstd: Standard deviation of heart rate [1 bpm] | 0.83 $\frac{\Delta\lambda}{\Delta hrstd}$= 72.37x10$^{-5}$ |
| lfhfratio: LF/HF ratio [0.5] | 0.16 |
| $\theta_i$ = angles of extrema [-70 -15 0 15 100] degrees | 0.98 $\frac{\Delta\lambda}{\Delta\theta_i(S_{BP})}$=3.02x10$^{-5}$ |
| BPoffset = Amplitude of blood pressure [80 mmHg] | -0.1061 |
| BPrange=Pulse pressure [40 mmHg] | -0.47 |

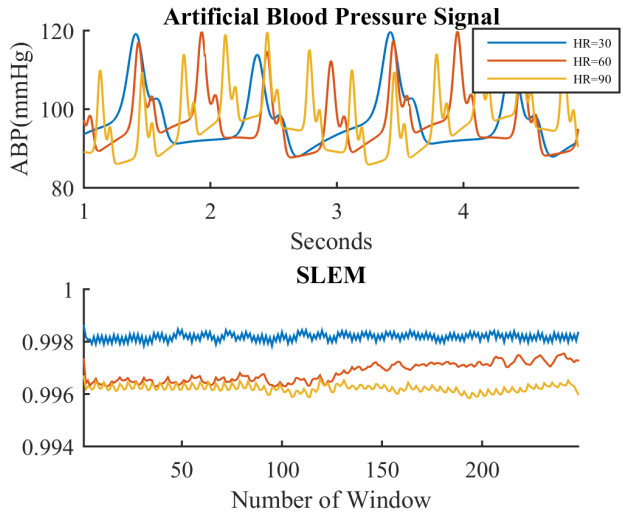

**Figure 8. Change in SLEM as the heart rate (HR) is changed for an artificial blood pressure model. Increasing heart rate causes a decrease in the SLEM.**

The results show that the change in the SLEM is explained mostly by an increase in heart rate, a decrease in heart rate variability and increase in frequency (in our case frequency of blood pressure was fixed at 100Hz). We also observed the effect of morphological change. It is shown in simulated hemorrhage with lower body negative pressure, that the time between two specific peaks of a blood pressure signal, defined as T14 parameter of Pulse Decomposition Analysis (PDA), decreases with progressive hemorrhage [16]. The fourth component ($S_{BP}$) of the parameter $\theta_i$ from the artificial blood pressure model was changed to simulate this change. From the results, we conclude,

$$HR\uparrow, HRV\downarrow, \theta_i(S_{BP})\uparrow \implies SLEM\downarrow$$

All of these trends in the parameters, that we observed as indicators of the decrease in the SLEM, have been described in the literature to change during hemorrhage [16] [14]. However, empirical data suggests different parameters in the blood pressure model induces competing changes in the SLEM (increase or decrease). The net effect is a contribution of the change in multiple signal parameters. But the combination of the parameters' change does not explain the change in SLEM adequately (the change in the SLEM is predicted to be much more in magnitude) if the net effect of the change in each parameter changes are considered. This might suggest that the change in the SLEM provides different information than captured by the signal parameters.

## 9. CONCLUDING DISCUSSION

We summarize, the change in the SLEM is explained mostly by the structural change in the transition probability matrix. This change indicates different information not captured by other existing approaches that have been studied. In the case of hemorrhage detection, the method works fairly well compared to other approaches to detect dynamical change. For high specificity, change in the SLEM could be used in conjunction with other parameters such as low blood pressure and increase in heart rate variability. For this particular application, hierarchical algorithm could be used. The first level of hemorrhage detection being the change in the SLEM with subsequent levels detecting hemorrhage in conjunction with change in other vital signs such as drop in blood pressure and increase in heart rate.

This index has shown to have high correlation with shock index (median correlation coefficient of -0.95). Shock index is used in clinical setting for evaluating the status of a patient due to hemodynamic instability [7]. It is defined as the ratio between heart rate and blood pressure. As hemorrhage progresses, due to the compensation mechanism between the sympathetic and the parasympathetic nervous system, blood pressure decreases, and to compensate that decrease for adequate tissue perfusion, heart rate increases; which is captured by the shock index. The change in the SLEM to have high correlation with shock index is of much significance in that regard. This indicates the hemodynamic imbalance is nicely captured within the signal of arterial blood pressure, requiring only one vital sign to be monitored.

The meaning of the change in the SLEM with regards to the states of a time series is another interesting research question and needs to be explored further. Our assumption is that, with progressive hemorrhage, as the pulse pressure (difference between systolic and diastolic blood pressure), the range of the states become smaller. At the same time, there is more variability in the state transitions which implies the system is creating a new evolution rule. As a consequence of both of these, more states are transitioning from one state to the other, which is reflected in the structural properties of the transition probability matrix.

It is also a valid argument that, if the change in the SLEM is mostly explained by the structural properties of the transition probability matrix, one can also use those properties (density or stationary probability) as the indicator of dynamical change of time series of complex systems. The counter to this argument is that, the change in the SLEM is the effect of the change in both of these parameters and hence is a stronger index. However, those structural properties can still be used, and the result may not be significantly different than using the SLEM.

One important aspect of our approach is finding the parameters of the algorithm empirically. For example, for the case of hemorrhage detection, we used 20 seconds of blood pressure data for each window with a sampling rate of 100 Hz (2000 samples). This window size was selected from a potential pool of many other combinations empirically studied with the FDA Meadow cluster after generating thousands of images. The change in the SLEM was best observed for this particular window size. It is possible that, to observe the dynamical change observed as the change in the SLEM, different applications would require different parameter settings for the algorithm.

One of the limitations of this study is that the algorithm was developed and studied for only one application, detection of hemorrhage. The method needs to be investigated further for other applications where dynamical change is expected.

## 10. ACKNOWLEDGMENT

We thank Dr. George Kramer from University of Texas Medical Branch (UTMB) at Galveston for providing the data. We are also grateful to Dr. Christopher Scully of US Food and Drug Administration (US FDA) and Dr. Richard Povinelli of Marquette University for their valuable contributions.